\documentclass[a4paper,12pt]{article}
\usepackage[warn]{mathtext}
\usepackage[T2A]{fontenc}
\usepackage[utf8]{inputenc}
\usepackage[english,russian]{babel}
\usepackage{amssymb,amsfonts,amsmath,mathtext,cite,enumerate,float}
\usepackage{graphicx} 
\usepackage{indentfirst}

\newcommand{\tocsecindent}{\hspace{6mm}}

\usepackage{geometry}
\makeatletter
\renewcommand{\@biblabel}[1]{#1.}
\makeatother

\begin{document}

\title{Влияние электронов на стабильность одномерных цепочек металлов}
\author{В.Д. Борман, П.В. Борисюк, О.С. Васильев, В.Н. Тронин, И.В. Тронин, \\М.А. Пушкин, В.И. Троян}
\date{}
\maketitle
\begin{abstract}
The physical model describing the influence of the electronic subsystem on the properties of one-dimensional chains of metal is presented. It is shown that depending on an interaction potential between atoms in one-dimensional system formation of chains of various length is possible. In case the characteristic depth of the potential well of the interatomic interaction does not exceed a certain magnitude, the chains in 1D system are formed with length of several angstroms, while the increase the depth of the well also leads to the possibility of formation of metal chains of greater length.\\

Представлена физическая модель, описывающая влияние электронной подсистемы на свойства одномерных цепочек металлов. Показано, что в зависимости от потенциала взаимодействия между атомами в одномерной системе возможно образование цепочек различной длины. В том случае, если характерная глубина потенциальной ямы межатомного взаимодействия не превышает определенной величины, в 1D-системе образуются цепочки с характерной длиной порядка нескольких ангстрем, в то время как увеличение глубины ямы приводит также к возможности образования цепочек металлов большей длины.
\end{abstract}

\begin{figure}[H]
\center{\includegraphics[width=0.7\linewidth]{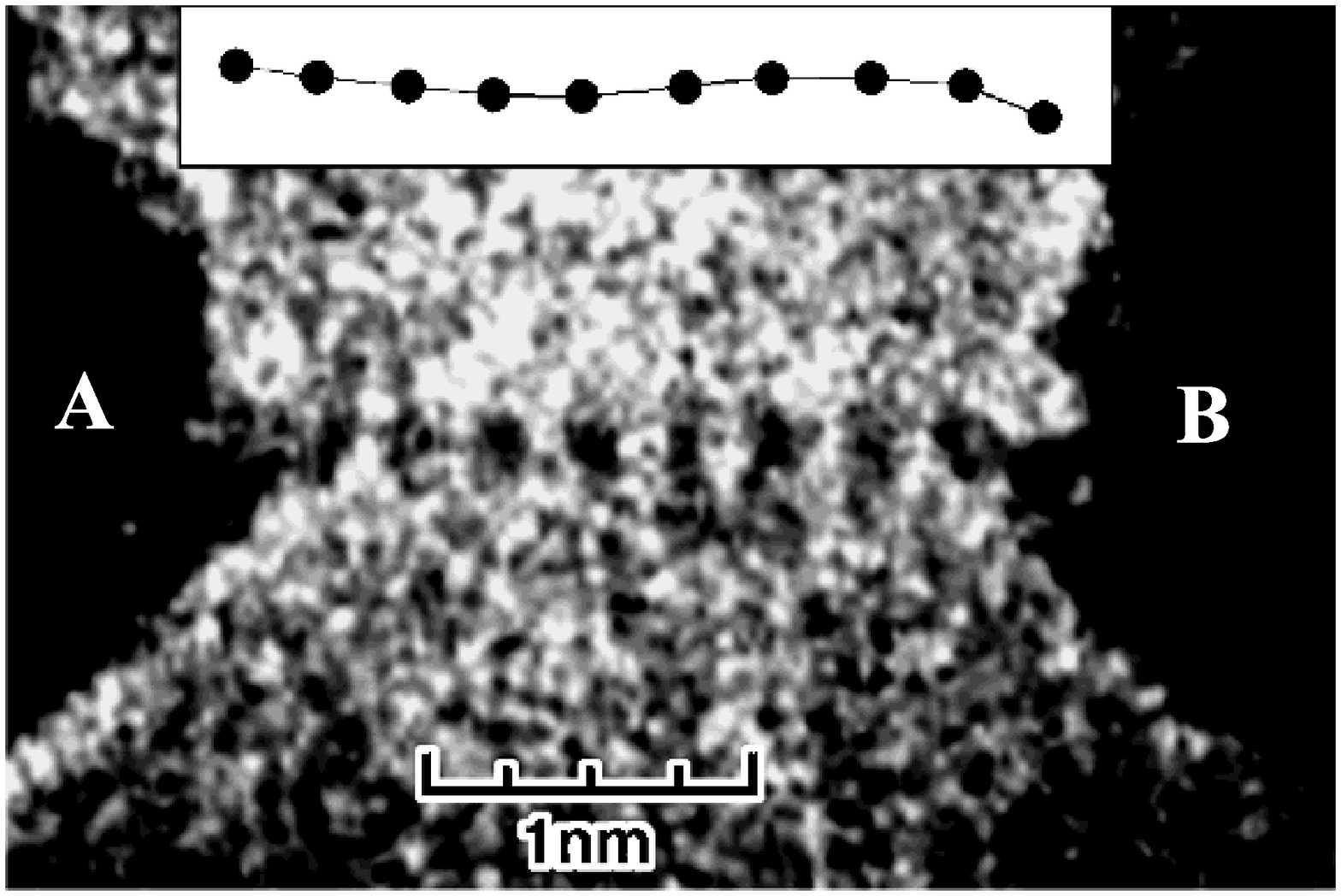}}
	\caption{\small{Изображение стабильной одномерной золотой атомной нити длиной 2.6 нм. Темные регионы по бокам изображения являются двумя иглами СТМ(А и В) \cite{Kizuka:2001wj}}}
	\label{fig:chain}
\end{figure}

Правильное понимание свойств наноразмерных контактов имеет решающее значение для многих областей современной нанотехнологии. Уникальные свойства моноатомных цепочек металлов привлекают в настоящее время значительное внимание как экспериментально\cite{Agrait:2003kr, Smit:2001gk, Csonka:2006uu, Kizuka:2001wj, Ohnishi:1998tz, Rodrigues:2001wv, RubioBollinger:2001us, Untiedt:2002iy, Yanson:1998uo}, так и теоретически \cite{Skorodumova:2005dg,Skorodumova:2003vj,Skorodumova:2000uc}. Атомарные цепочки могут быть получены в экспериментах по механически контролируемому обрыву цепи с использованием туннельного микроскопа или просвечивающего электронного микроскопа\cite{Agrait:2003kr}. Структуры, получаемые в данных экспериментах являются одномерными цепочками, состоящие из нескольких атомов металла, находящиеся между двух поверхностей(см рис.~\ref{fig:chain}).

\begin{figure}[H]
\begin{minipage}[h]{0.49\linewidth}
\center{\includegraphics[width=0.95\linewidth]{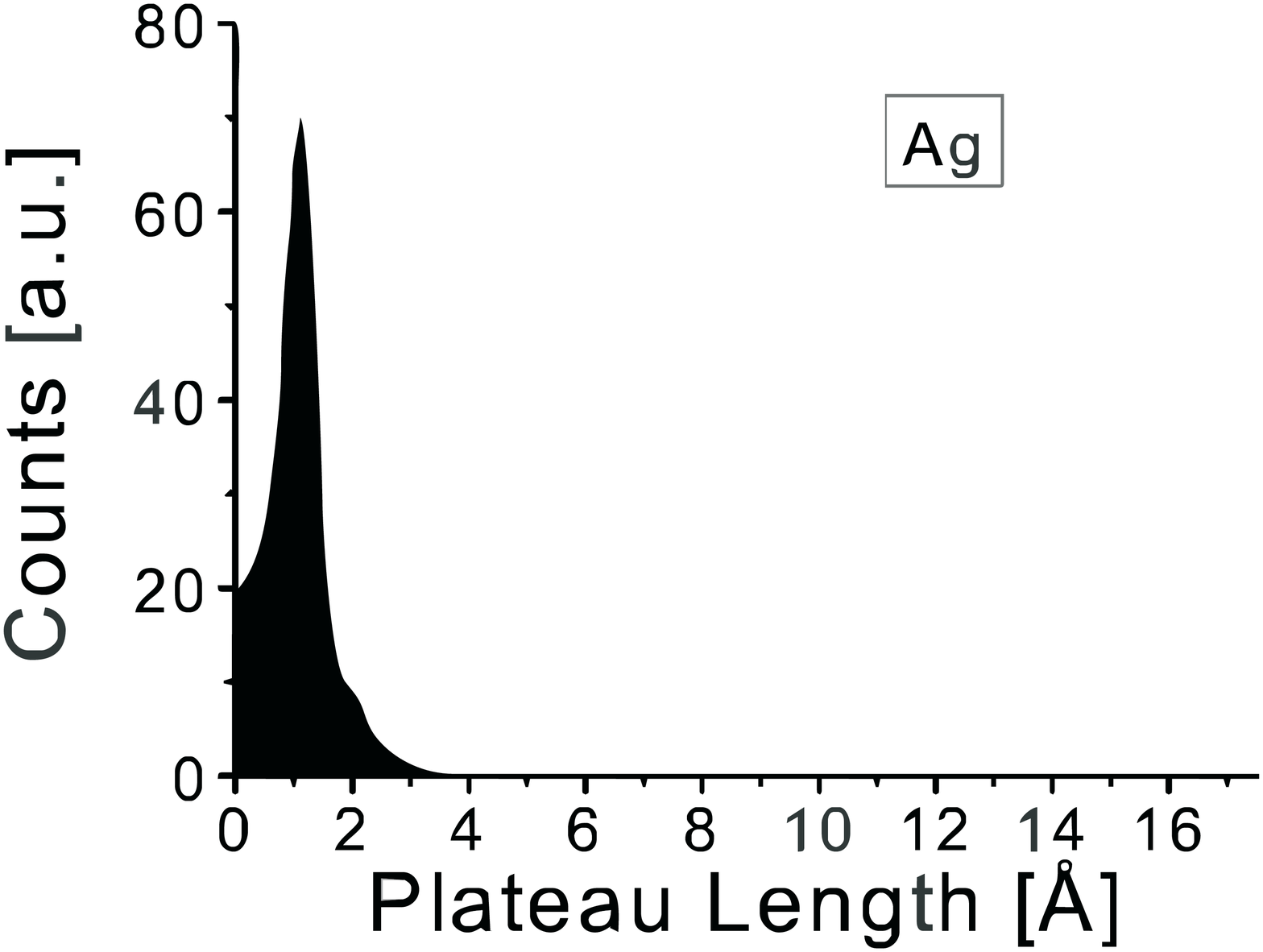}}
\end{minipage}
\hfill
\begin{minipage}[h]{0.49\linewidth}
\center{\includegraphics[width=0.95\linewidth]{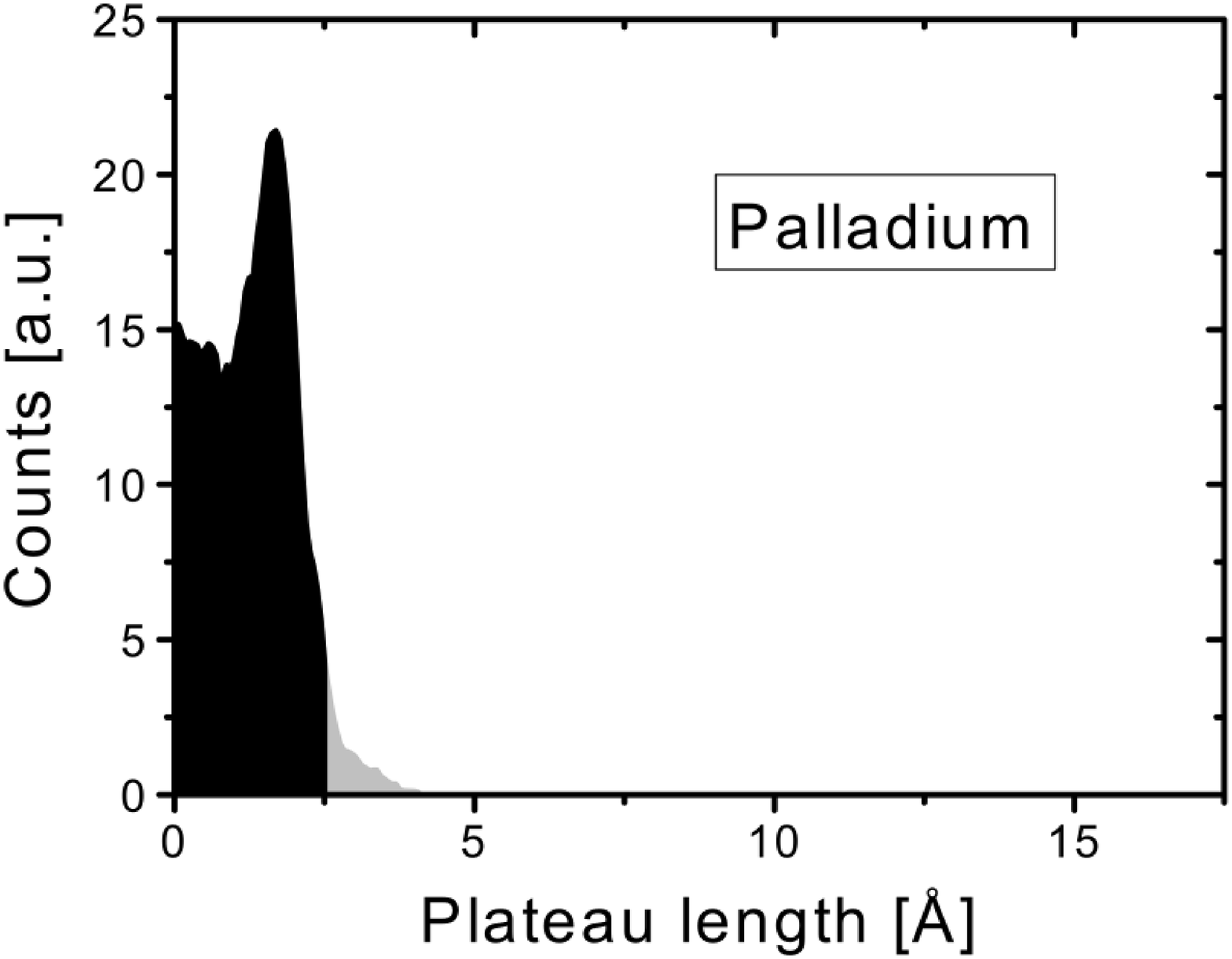}}
\end{minipage}
\vfill
\begin{minipage}[h]{0.49\linewidth}
\center{\includegraphics[width=0.95\linewidth]{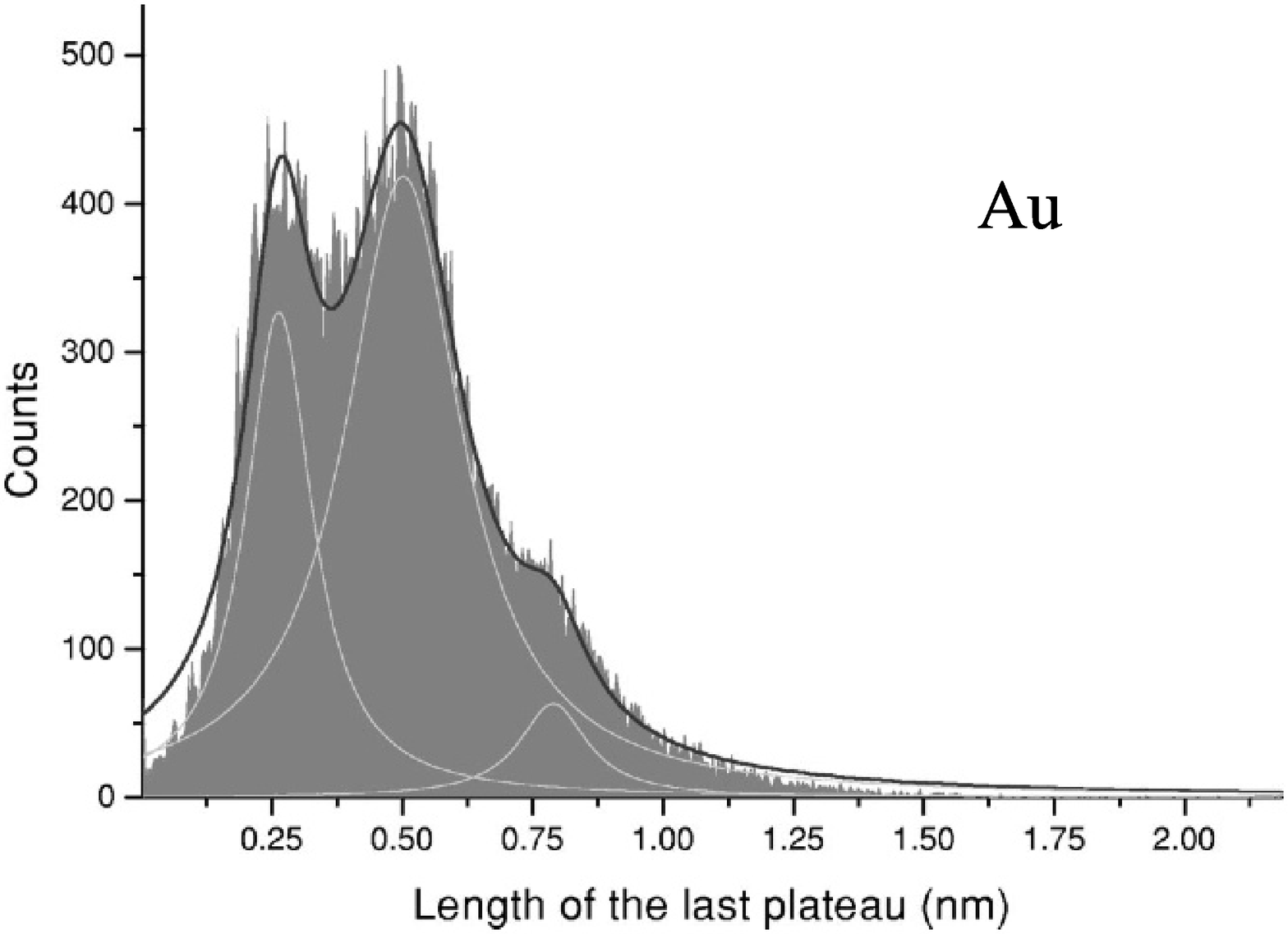}}
\end{minipage}
\hfill
\begin{minipage}[h]{0.49\linewidth}
\center{\includegraphics[width=0.95\linewidth]{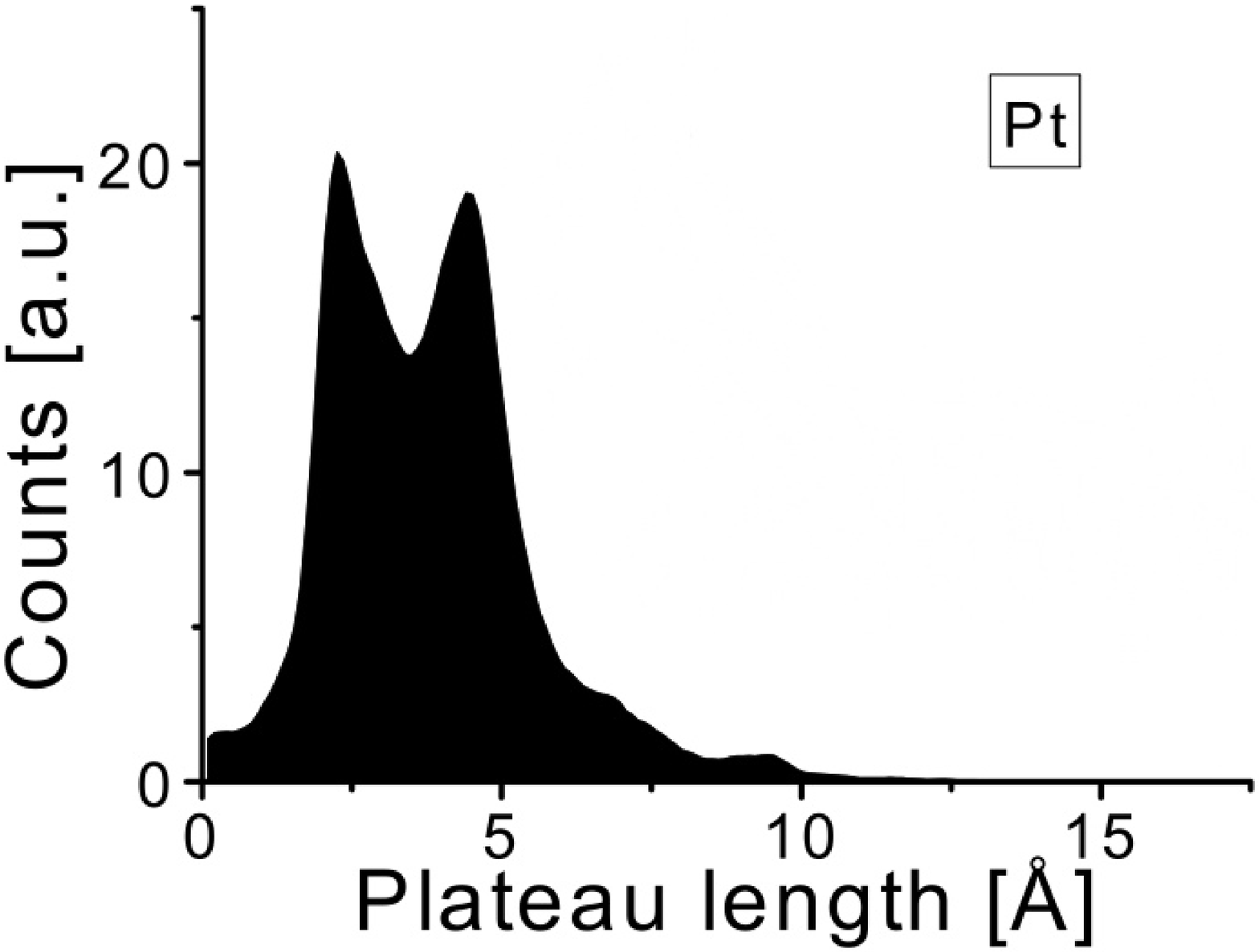}}
\end{minipage}
	\caption{\small{Histogram of lengths for the last conductance plateau для  Ag, Pd, Pt\cite{Smit:2001gk}, Au\cite{Untiedt:2002iy},}}
	\label{fig:agaupt}
\end{figure}

Образование подобных цепочек сильно зависит от материала атомов, из которых состоит цепочка. Было показано, что золотые цепочки могут быть до 2.6 нм в длину\cite{Untiedt:2002iy,Yanson:1998uo}, тогда как цепочки из атомов серебра в длину не превышают нескольких ангстрем\cite{Smit:2001gk}. Так, в работах\cite{Smit:2001gk,Untiedt:2002iy} с помощью методики механически контролируемого обрыва цепи (mechanically controllable break-junction (MCB)),в серии экспериментов были получены гистрограммы, отображающие частоту обрыва цепочек металлов Ag, Au, Pt, Pd от длины цепочки(см. рис.~\ref{fig:agaupt}). 

Таким образом, можно сделать вывод, что металлы Au и Pt, в отличие от Ag и Pd, могут образовывать одномерные цепочки с количеством атомов $N>2$. В связи с отсутствием теоретического описания данного эффекта в настоящее время, необходимо проведение исследования свойств одномерных наноцепочек. Понимание механизмов образования цепочек может помочь улучшить контроль за процессом их получения, что, в свою очередь, может привести к созданию цепочек различных материалов с уникальными свойствами (магнетизм, сверхпроводимость) или созданию цепочек больших размеров. 

В данной работе представлена физическая модель, описывающая влияние электронной подсистемы на свойства одномерных цепочек металлов. Показано, что в зависимости от потенциала взаимодействия между атомами в одномерной системе возможно образование цепочек различной длины. В том случае, если характерная глубина потенциальной ямы межатомного взаимодействия не превышает определенной величины, в 1D-системе образуются цепочки с характерной длиной порядка нескольких ангстрем, в то время как увеличение глубины ямы приводит также к возможности образования цепочек металлов большей длины.

Рассмотрим влияние электронной подсистемы на свойства одномерных цепочек металлов. Предположим, что цепочки металлов, наблюдаемые в экспериментах, являются реализацией возможных состояний одномерной статистической системы. Рассмотрим бесконечную одномерную цепочку. Рассмотрим электрон в поле флуктуаций плотности одномерной системы частиц. Пусть $\overline{n}$~---~средняя плотность частиц в рассматриваемой системе, а $n=n(x)=\sum_i \delta (x-x_i)$~---~микроскопическая плотность. Тогда $\delta n = \overline{n}-n$~---~флуктуация плотности. Гамильтониан электрона в поле флуктуаций плотности будет иметь вид:

\begin{equation}
	\label{eq:math:H1}
	H = H[\Psi ,\delta n] + H[\delta n]
\end{equation}

где первое слагаемое \eqref{eq:math:H1} --- часть гамильтониана, отвечающая взаимодействию электрона с флуктуацией плотности, а второе слагаемое \eqref{eq:math:H1}, отвечающее собственно флуктуации плотности, представляет собой изменение свободной энергии системы частиц:

\begin{equation}
	\label{eq:math:Hpsin}
	H[\Psi ,\delta n] = \int \left( - \Psi ^* \frac{{\hbar^2 }}{{2m}}(\Delta \Psi ) + \Psi ^* U\Psi \right)\,dx \end{equation}

\begin{equation}
	\label{eq:math:Hn}
	\begin{split}
	H[\delta n] = F(n + \delta n) - F(n) = \int \left. \frac{\delta F}{\delta n} \right|_{\overline n} \delta n \,dx + \frac{1}{2} \int \frac{\delta^2 F}{\delta n(x) \delta n(x')} \, \delta n(x) \, \delta n(x') \, dx dx' ={} \\
	{} = - \frac{1}{2}\int {\beta ^{ - 1} (x,x')\delta n(x)\delta n(x')\,dxdx'} 
	\end{split}
\end{equation}

где $\beta$  --- функция отклика системы, $\beta^{-1}=\frac{\delta^2 F}{\delta n(x) \delta n(x')}$, $F$ --- свободная энергия системы. 
Потенциальная энергия электрона в поле флуктуаций плотности $U=\int{V(x-x')\, dx\, \delta n}$. Будем предполагать одночастичный потенциал $V$ локальным: $V(x-x') = V_0 \delta(x-x')$, тогда $U = V_0 \delta n(x)$. Это приближение, принимаемое при изучении рассеяния медленных частиц, справедливо, когда де-Бройлевская длина волны электрона много больше размера рассеивателя. Это условие выполняется, поскольку длина волны электрона порядка размера кластера, а кластер содержит много атомов($N \gg 1$).\\

Для равновесного состояния:

\begin{equation}
	\label{eq:math:dHdn}
	\frac{\delta H}{\delta (\delta n)}=0
\end{equation}

Подставляя \eqref{eq:math:H1} в \eqref{eq:math:dHdn}, получим 

\begin{equation}
	\label{eq:math:dHdn1}
	\begin{split}
	\int{V_0\Psi^*(x)\Psi(x)\,dx} &-\int{\beta^{-1}(x,x')\delta n'(x')\,dxdx'}=0; \\
	 {}   \delta n'(x')&=\int{\beta(x,x') V_0 \Psi^*(x) \Psi(x)\,dx}
	\end{split}
\end{equation}

Таким образом, из \eqref{eq:math:Hpsin}, \eqref{eq:math:Hn}, с учетом \eqref{eq:math:dHdn1} получаем:

\begin{equation}
	\begin{split}
	H_{eff} [\Psi ] =\left. H[\Psi]\right|_{\delta n'} = \int \left( - \Psi ^* \frac{{h^2 }}{{2m}}\Delta \Psi \right)\,dx + \frac{1}{2}V_0 \int \beta(x,x') \left(\Psi ^*(x) \Psi(x) \right) \times {}\\
{} \times \left(\Psi ^*(x') \Psi(x') \right) \,dxdx'  
	\end{split}
 \label{eq:math:Heff}
\end{equation}

Функция отклика $\beta(x,x')$ связана с корреляционной функцией $\nu(x,x')$ флуктуационно-диссипативной теоремой \cite{balesku1978}:

\begin{equation}
	\label{eq:math:nu_beta}
	n^2 \nu(x,x')=-\delta (x-x') n - T \beta(x,x')
\end{equation}

или, в $k$ --- представлении, 

\begin{equation}
	\label{eq:math:nu_beta_k}
	\beta(k)=-\frac{n}{T}[1+n\nu (k)]
\end{equation}

Для упрощения воспользуемся локальным приближением (что соответствует длинноволновому приближению($k \rightarrow 0$)) \cite{Devyatko1990a} :

\begin{equation}
	\label{eq:math:nu_beta_k1}
	\begin{split}
	\beta(k)=&-\frac{n}{T}[1+n\nu(k)+n\nu_{cor}(k)-n\nu_{cor}(k)]=-\frac{n}{T}[1+n\nu_{cor}(k)]- {}\\
	{}&- \frac{n}{T}[n\nu(k)-n\nu_{cor}(k)] \approx  -\frac{n}{T}[1+n\nu_{cor}(0)]-\frac{n}{T}[n\nu(0)-n\nu_{cor}(0)]\delta(k)\frac{1}{a}
	\end{split}
\end{equation}

здесь и далее индекс ${cor}$ --- соответствует величинам, вычисленным для системы частиц с взаимодействием типа "`твердые шары"', величины без индекса соответствуют вычисленным для системы частиц с полным взаимодействием.
Функция отклика $\beta_{k=0}$ связана с сжимаемостью соотношением \cite{Korneev}:

\begin{equation}
\label{eq:math:beta0}
\left. \frac{\partial p}{\partial n} \right|_{cor}=n \beta^{-1}_{k=0}=-T [1+n \nu_{cor}(0)]^{-1}
\end{equation}

в \eqref{eq:math:beta0} перейдем от плостности $n$ к степени заполнения $\theta$, $\theta=n a$ --- степень заполнения в одномерной системе: 

\begin{equation}
\label{eq:math:beta1}
\frac{1}{a} \left. \frac{\partial p}{\partial \theta}\right|_{cor}=-T[1+n\nu_{cor}(0)]^{-1}
\end{equation}

Сжимаемость $\left. \frac{\partial \theta}{\partial p}\right|_{cor}$, где $p$ --- одномерное давление, для взаимодействия типа "`твердые шары"' может быть получена из уравнения состояния одномерного газа, которое может быть найдено из cоотношений \cite{fisher1961}:

\begin{equation}
\label{eq:math:sost1d}
\begin{split}
\frac{1}{n} +& T \frac{\partial}{\partial p} \ln(\varphi(p,T))=0 \\
\varphi(p,T) =& \int \limits^{\infty}_0 {dx \,\exp \left(-\frac{px+\Phi(x)}{T}\right)}
\end{split}
\end{equation}

где $\Phi(x)$ --- потернциал взаимодействия между атомами. Таким образом, уравнение состояния $p(\theta,T)$ и восприимчивость $\left. \frac{\partial \theta}{\partial p}\right|_{cor}$ одномерного газа частиц для взаимодействия типа "`твердые шары"' будет иметь вид:

\begin{equation}
\label{eq:math:p1d}
p = \frac{T}{a}\frac{\theta }{{1 - \theta }}, \quad \left. \frac{\partial \theta}{\partial p}\right|_{cor}=\frac{(1-\theta)^2}{T} a
\end{equation}

где $a$ --- радиус частицы,   из \eqref{eq:math:nu_beta_k} и \eqref{eq:math:beta1} и \eqref{eq:math:p1d}  для $\beta(k)$ получаем:

\begin{equation}
	\label{eq:math:betaK}
	\beta (k) = \frac{\theta }{aT}\left( {1 - \theta } \right)^2  - \frac{\theta }{{Ta^3 }}\left( {T\frac{{\partial \theta }}{{\partial p}} - a\left( {1 - \theta } \right)^2 } \right)\delta (k)
\end{equation}

Восприимчивость системы взаимодействующих частиц $\frac{{\partial \theta }}{{\partial p}}$ найдем в приближении взаимодействия типа "`прямоугольная яма"':

\begin{equation}
\label{eq:math:yama}
\Phi \left( x \right) = \left\{ {\begin{array}{*{20}c}
   {\infty ,x \le a }  \\
   { - \varepsilon, a  < x \le a  + R}  \\
   {0,x > a  + R}  \\
\end{array}} \right.
\end{equation}

Здесь $\varepsilon$ --- глубина ямы, $a$ --- радиус частицы $R$ --- ширина ямы. В этом случае восприимчивость $\frac{{\partial \theta }}{{\partial p}}$ может быть найдена из уравнения состояния для одномерного газа \cite{herzfeld1932}:

\begin{equation}
   \label{eq:math:p1d_teta}
   p_{1D} \sigma \left( {\frac{1}{\theta } - 1} \right) = T - p_{1D} R\left[ {\frac{{\exp \left( {\beta p_{1D} R} \right)}}{{1 - \exp \left( { - \varepsilon/T} \right)}} - 1} \right]^{ - 1} 
\end{equation}

где $p_{1D}=\frac{pa}{T}$ --- безразмерное одномерное давление. Таким образом, с учетом \eqref{eq:math:betaK}, из \eqref{eq:math:Heff} для эффективного гамильтониана $H_{eff}$ получим:

\begin{equation}
	\label{eq:math:Heff1}
	H_{eff} [\Psi ] = \int {\left( { - \Psi ^* \frac{{\hbar ^2 }}{{2m}}\Delta \Psi } \right)dx - \frac{{V_0 ^2 }}{{4\pi }}} \frac{\lambda }{T}\int {\left| \Psi  \right|} ^4 dx - \frac{{V_0 ^2 }}{{4\pi }}\frac{\alpha }{T}\left( {\int {\left| \Psi  \right|} ^2 dx} \right)^2 
\end{equation}

Здесь $\lambda  = \frac{{\theta (1 - \theta )^2 }}{a}$, $\alpha  = \frac{\theta }{{a^3 }}\left[ {T\frac{{\partial \theta }}{{\partial p}} - a(1 - \theta )^2 } \right]$

Зная эффективный гамильтониан электрона в поле флуктуации плотности \eqref{eq:math:Heff1}, получим уравнение Шредингера:

\begin{equation}
	\label{eq:math:nelin_shed}
	\Delta \Psi  =  - \frac{m}{{\hbar ^2 }}\frac{{V_0^2 \lambda }}{T}\Psi ^3  - \left( {\frac{m}{{\hbar ^2 }}\frac{{V_0^2 \alpha }}{{2\pi T}} + E\frac{m}{\hbar ^2 }} \right)\Psi 
\end{equation}

Для такого уравнения есть термин "`нелинейное уравнение Шредингера"'. Показано~\cite{alimenkov2006}, что одним из решений подобного нелинейного уравнения будет является решение вида $\Psi(x)=A/{\mathrm{ ch}}(Bx)$, где $A$ и $B$ --- константы, $B$ имеет смысл обратного характерного размера солитона. Таким образом, в нашем случае решением будет являться:

\begin{equation}
	\label{eq:math:psi_resh}
	\Psi (x) = \frac{V_0}{\hbar}\sqrt{\frac{\sqrt{2\pi}m}{8Ta}\theta (1-\theta)^2}\,\frac{1}{{{\mathrm{ch}}\left( {\frac{\sqrt{2\pi}}{4}\frac{mV_0^2}{\hbar^2 T}\frac{\theta}{a}(1-\theta)^2\,x} \right)}}
\end{equation}	

где амплитуда солитона $A$:

\begin{equation}
\label{eq:A}
A=\frac{V_0}{\hbar}\sqrt{\frac{\sqrt{2\pi}m}{8Ta}\theta (1-\theta)^2}
\end{equation}

а обратный характерный размер солитона $B$:

\begin{equation}
\label{eq:B}
B=\frac{\sqrt{2\pi}}{4}\frac{mV_0^2}{\hbar^2 T}\frac{\theta}{a}(1-\theta)^2;
\end{equation}

Заметим, что зависимость \eqref{eq:B} имеет максимальное значение при $\theta=1/3$, что соответствует минимальной длине цепочки $L_{min}=1/B=\sqrt{2\pi}\frac{mV_0^2}{\hbar^2 Ta}\frac{1}{27}$.

Зная волновую функцию электрона, можно найти энергию электрона в зависимости от степени заполнения $\theta$ одномерной системы частиц. Из \eqref{eq:math:psi_resh} получим, что энергия электрона будет иметь вид:

\begin{equation}
\label{eq:E_sol_f}
E=-\sqrt{\frac{\pi}{2}}\frac{V_0^2}{Ta^3}\,\theta\left(T\frac{\partial \theta}{\partial p} - a(1-\theta)^2\right)-\frac{\pi}{16}\frac{mV_0^4}{\hbar^2 T^2a^2}\,\theta^2(1-\theta)^4,
\end{equation}

Величину $\frac{\partial \theta}{\partial p}$ в приближении взаимодействия типа ”прямоугольная яма“ можно получить из \eqref{eq:math:p1d}. В таком случае, зависимость энергии электрона $E$ от степени заполнения одномерной системы $\theta$ при различных значениях глубины потенциальной ямы взаимодействия атомов представлена на рис.~\ref{fig:energ(t)}.

\begin{figure}[H]
\begin{minipage}[h]{0.49\linewidth}
\input{Etet_2.tex}	
\end{minipage}
\hfill
\begin{minipage}[h]{0.49\linewidth}
\input{Etet.tex}	
\end{minipage}
\caption{\small{Зависимость энергии электрона от степени заполнения одномерной системы $\theta$ при различных значениях глубины потенциальной ямы взаимодействия атомов $\varepsilon=5\cdot 10^{-2}$~эВ~(а) и $\varepsilon=9\cdot 10^{-2}$~эВ~(б)}}
	\label{fig:energ(t)}
\end{figure}

Как можно видеть из рис.~\ref{fig:energ(t)}, при малой глубине потенциальной ямы взаимодействия атомов $\varepsilon=5\cdot 10^{-2}$ эВ, зависимость энергии электрона от степени заполнения имеет один минимум при $\theta \simeq 0.22$ (см. рис.~\ref{fig:energ(t)}(a)), что отвечает одному наиболее вероятному состоянию электрона при любом значении величины степени заполнения $\theta$. При увеличении же глубины потенциальной ямы $\varepsilon=9\cdot 10^{-2}$ эВ, в зависимости энергии электрона от степени заполнения появляется второй минимум при $\theta \simeq 0.68$ (см. рис.~\ref{fig:energ(t)}(a)), таким образом при увеличении энергии взаимодействия атомов имеется два наиболее вероятных состояний электрона, соответствующие различным $\theta$.

\begin{figure}[H]
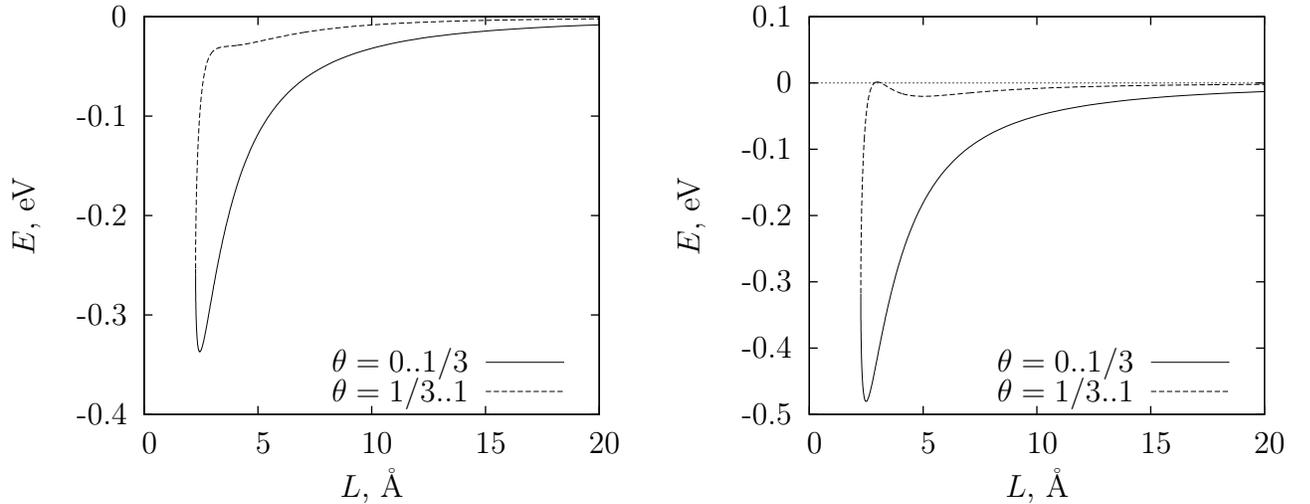

\begin{minipage}[h]{0.49\linewidth}
\input{EL_2.tex}	
\end{minipage}
\hfill
\begin{minipage}[h]{0.49\linewidth}
\input{EL.tex}	
\end{minipage}
\caption{\small{Зависимость энергии электрона от характерной длины солитона $L$  при различных значениях глубины потенциальной ямы взаимодействия атомов $\varepsilon=5\cdot 10^{-2}$~эВ~(а) и $\varepsilon=9\cdot 10^{-2}$~эВ~(б)}}
	\label{fig:energ(l)}
\end{figure}

Зная связь характерной длины солитона $1/B$ со степенью заполнения одномерной системы $\theta$ \eqref{eq:B}, из \eqref{eq:E_sol_f} можно найти зависимость энергии электрона от характерной длины солитона $E(1/B)=E(L)$ (см. рис.~\ref{fig:energ(l)}). В связи с  громоздкостью, итоговое аналитическое выражение для $E(L)$ не приводится.

Как можно видеть из рис.~\ref{fig:energ(l)}(б), при глубине потенциальной ямы взаимодействия атомов $\varepsilon=9\cdot 10^{-2}$ эВ имеются два минимума энергии электрона при $L\simeq 4.9, 8$ \AA, соответствующие различным диапазонам $\theta$. Подобное поведение можно объяснить следующим образом. При $\theta=0\div 1/3$ наиболее вероятным является образование цепочек длиной $L \simeq 4.9$ \AA. С увеличением же степени заполнения $\theta > 1/3$ система переходит в новое состояние с образованием цепочек длиной $L\simeq 8$ \AA. 

Получим вероятность образования цепочек длинной $L$. Величина $\exp (-E(L)/T)$ --- есть вклад в плотность вероятности образования цепочки с характерным размером $L$  за счет взаимодействия электрона с флуктуацией плотности. Для того, чтобы получить вероятность образования цепочки длиной $L$ необходимо дополнительно учесть вероятность образования флуктуации плотности в одномерной системе. Таким образом, плотность вероятности образования цепочки с характерным размером $L$ будет иметь вид:

\begin{equation}
\label{eq:f_ver}
f=\exp \left(\frac{-E(L)}{T}\right)w_L
\end{equation}

где $w_L$ --- вероятность найти в одномерной системе невзаимодействующих частиц цепочку длиной $L$. $w_L$ определим следующим образом. Известно \cite{landau5}, что вероятность того, что на длине $L$ будет находиться всего $N$ атомов дается формулой Пуассона:

\begin{equation}
\label{fpuassona}
w_N=\frac{\overline{n}^N \exp{(-\overline{n})}}{N!},
\end{equation}

где $\overline{n}=\frac{N_0}{L_0}L=\frac{\theta}{2a}L$ --- среднее значение числа частиц на длине $L$, $N_0$ --- число частиц в системе, $L_0$ --- длина системы, $N$ --- число частиц в цепочке. Полагая, что в цепочке длиной $L$ содержится $N$ частиц при плотности частиц $\theta'=1$, получим  $N=\frac{L}{2a}$. Таким образом, вероятность флуктуационного образования цепочки из $N$ частиц будет выглядеть следующим образом:

\begin{equation}
\label{fpuassona_N}
w_N=\frac{(\theta N)^N \exp{(-(\theta N))}}{N!}
\end{equation}

Имея в виду вышеозначенную связь числа частиц с длиной цепочки, получим:

\begin{equation}
\label{fpuassona_L}
w_L=\frac{\left( \frac{\theta L}{2a}\right)^{\frac{L}{2a}} \exp{\left(- \frac{\theta L}{2a}\right)}}{\left(\frac{L}{2a}\right)!}
\end{equation}

Таким образом, плотность вероятности образования цепочки длиной $L$ в одномерной системе будет выглядеть следующим образом:

\begin{equation}
\label{eq:f_ver_full}
f=\exp \left(\frac{-E(L)}{T}\right) \frac{\left( \frac{\theta(L) L}{2a}\right)^{\frac{L}{2a}} \exp{\left(- \frac{\theta(L) L}{2a}\right)}}{\left(\frac{L}{2a}\right)!},
\end{equation}

где зависимость $\theta(L)=\theta(1/B)$ описывается выражением \eqref{eq:B}.

\begin{figure}[H]
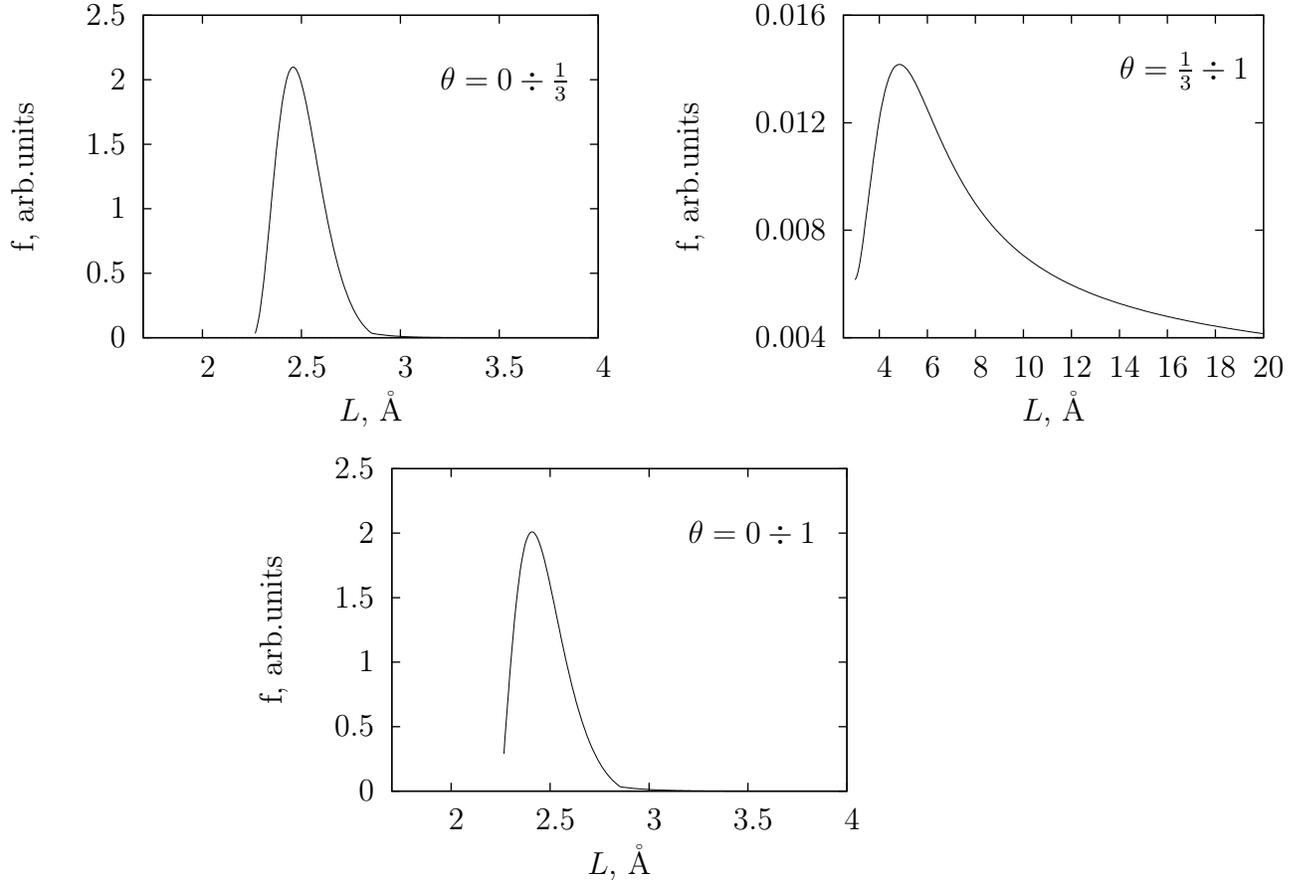

\begin{minipage}[h]{0.49\linewidth}
\input{frasp0_13.tex}	
\end{minipage}
\hfill
\begin{minipage}[h]{0.49\linewidth}
\input{frasp13_1.tex}	
\end{minipage}
\vfill
\begin{minipage}[h]{0.9\linewidth}
\begin{center}
\input{frasp_2.tex}	
\end{center}
\end{minipage}
	\caption{\small{Зависимость плотности функции распределения от длины цепочки $L$ при значении глубины потенциальной ямы взаимодействия атомов $\varepsilon=9\cdot 10^{-2}$~эВ, $\theta=0\div \frac{1}{3}$(а), $\theta=\frac{1}{3}\div 1$(б), при $\varepsilon=5\cdot 10^{-2}$~эВ,$\theta=0\div 1$(в) }}
	\label{fig:frasp(l)}
\end{figure}

На рис.~\ref{fig:frasp(l)} представлены зависимости плотности вероятности образования цепочек длиной $L$ при глубине потенциальной ямы взаимодействия атомов $\varepsilon=9\cdot 10^{-2}$~эВ (рис.~\ref{fig:frasp(l)}(а), (б)) и $\varepsilon=5\cdot 10^{-2}$(в). 

Как можно видеть из рис.~\ref{fig:frasp(l)}, при значении глубины потенциальной ямы взаимодействия атомов $\varepsilon=5\cdot 10^{-2}$~эВ, при любой плотности $\theta$ в одномерной системе наиболее вероятным является образование димеров - цепочек, состоящих из двух атомов, тогда как при увеличении глубины ямы появляется дополнительный пик плотности функции распределения, отвечающий образованию цепочек с $N>2$.

Таким образом, показано, что в зависимости от потенциала взаимодействия между атомами в одномерной системе возможно образование цепочек различной длины. В том случае, если характерная глубина потенциальной ямы межатомного взаимодействия не превышает определенной величины, в 1D-системе образуются цепочки с характерной длиной порядка нескольких ангстрем, в то время как увеличение глубины ямы приводит к возможности образования цепочек металлов также и большей длины.\\

Работа выполнена при финансовой поддержке Федеральной целевой программы "Научные и научно-педагогические кадры инновационной России на 2009 – 2013 годы".

\addcontentsline{toc}{chapter}{\tocsecindent{Литература}}
\bibliography{biblio1} 
\end{document}